

%

%
 \font\twelvebf=cmbx12
 \font\twelvett=cmtt12
 \font\twelveit=cmti12
 \font\twelvesl=cmsl12
 \font\twelverm=cmr12		\font\ninerm=cmr9
 \font\twelvei=cmmi12		\font\ninei=cmmi9
 \font\twelvesy=cmsy10 at 12pt	\font\ninesy=cmsy9
 \skewchar\twelvei='177		\skewchar\ninei='177
 \skewchar\seveni='177	 	\skewchar\fivei='177
 \skewchar\twelvesy='60		\skewchar\ninesy='60
 \skewchar\sevensy='60		\skewchar\fivesy='60
%
%

%
 \font\fourteenrm=cmr12 scaled 1200
 \font\seventeenrm=cmr12 scaled 1440
 \font\fourteenbf=cmbx12 scaled 1200
 \font\seventeenbf=cmbx12 scaled 1440
%
%

%
%
%
\font\tenmsb=msbm10
\font\twelvemsb=msbm10 scaled 1200
\newfam\msbfam

%
\font\tensc=cmcsc10
\font\twelvesc=cmcsc10 scaled 1200
\newfam\scfam

%
\def\seventeenpt{\def\rm{\fam0\seventeenrm}%
 \textfont\bffam=\seventeenbf	\def\bf{\fam\bffam\seventeenbf}}
\def\fourteenpt{\def\rm{\fam0\fourteenrm}%
 \textfont\bffam=\fourteenbf	\def\bf{\fam\bffam\fourteenbf}}
\def\twelvept{\def\rm{\fam0\twelverm}%
 \textfont0=\twelverm	\scriptfont0=\ninerm	\scriptscriptfont0=\sevenrm
 \textfont1=\twelvei	\scriptfont1=\ninei	\scriptscriptfont1=\seveni
 \textfont2=\twelvesy	\scriptfont2=\ninesy	\scriptscriptfont2=\sevensy
 \textfont3=\tenex	\scriptfont3=\tenex	\scriptscriptfont3=\tenex
 \textfont\itfam=\twelveit	\def\it{\fam\itfam\twelveit}%
 \textfont\slfam=\twelvesl	\def\sl{\fam\slfam\twelvesl}%
 \textfont\ttfam=\twelvett	\def\tt{\fam\ttfam\twelvett}%
 \scriptfont\bffam=\tenbf 	\scriptscriptfont\bffam=\sevenbf
 \textfont\bffam=\twelvebf	\def\bf{\fam\bffam\twelvebf}%
 \textfont\scfam=\twelvesc	\def\sc{\fam\scfam\twelvesc}%
 \textfont\msbfam=\twelvemsb	
 \baselineskip 14pt%
 \abovedisplayskip 7pt plus 3pt minus 1pt%
 \belowdisplayskip 7pt plus 3pt minus 1pt%
 \abovedisplayshortskip 0pt plus 3pt%
 \belowdisplayshortskip 4pt plus 3pt minus 1pt%
 \parskip 3pt plus 1.5pt
 \setbox\strutbox=\hbox{\vrule height 10pt depth 4pt width 0pt}}
\def\tenpt{\def\rm{\fam0\tenrm}%
 \textfont0=\tenrm	\scriptfont0=\sevenrm	\scriptscriptfont0=\fiverm
 \textfont1=\teni	\scriptfont1=\seveni	\scriptscriptfont1=\fivei
 \textfont2=\tensy	\scriptfont2=\sevensy	\scriptscriptfont2=\fivesy
 \textfont3=\tenex	\scriptfont3=\tenex	\scriptscriptfont3=\tenex
 \textfont\itfam=\tenit		\def\it{\fam\itfam\tenit}%
 \textfont\slfam=\tensl		\def\sl{\fam\slfam\tensl}%
 \textfont\ttfam=\tentt		\def\tt{\fam\ttfam\tentt}%
 \scriptfont\bffam=\sevenbf 	\scriptscriptfont\bffam=\fivebf
 \textfont\bffam=\tenbf		\def\bf{\fam\bffam\tenbf}%
 \textfont\scfam=\tensc		\def\sc{\fam\scfam\tensc}%
 \textfont\msbfam=\tenmsb	
 \baselineskip 12pt%
 \abovedisplayskip 6pt plus 3pt minus 1pt%
 \belowdisplayskip 6pt plus 3pt minus 1pt%
 \abovedisplayshortskip 0pt plus 3pt%
 \belowdisplayshortskip 4pt plus 3pt minus 1pt%
 \parskip 2pt plus 1pt
 \setbox\strutbox=\hbox{\vrule height 8.5pt depth 3.5pt width 0pt}}

%
\def\twelvepoint{%
 \def\small{\tenpt\rm}%
 \def\normal{\twelvept\rm}%
 \def\big{\fourteenpt\rm}%
 \def\huge{\seventeenpt\rm}%
 \footline{\hss\twelverm\folio\hss}
 \normal}
%

%
\def\bigbold{\big\bf}

%
\catcode`\@=11
%
%
\def\footnote#1{\edef\@sf{\spacefactor\the\spacefactor}#1\@sf
 \insert\footins\bgroup\small
 \interlinepenalty100	\let\par=\endgraf
 \leftskip=0pt		\rightskip=0pt
 \splittopskip=10pt plus 1pt minus 1pt	\floatingpenalty=20000
 \smallskip\item{#1}\bgroup\strut\aftergroup\@foot\let\next}
%
%
%
%
\def\hexnumber@#1{\ifcase#1 0\or 1\or 2\or 3\or 4\or 5\or 6\or 7\or 8\or
 9\or A\or B\or C\or D\or E\or F\fi}
\edef\msbfam@{\hexnumber@\msbfam}

%
%
%
\catcode`\@=12

\newcount\EQNO      \EQNO=0
\newcount\FIGNO     \FIGNO=0
\newcount\REFNO     \REFNO=0
\newcount\SECNO     \SECNO=0
\newcount\SUBSECNO  \SUBSECNO=0
\newcount\FOOTNO    \FOOTNO=0
\newbox\FIGBOX      \setbox\FIGBOX=\vbox{}
\newbox\REFBOX      \setbox\REFBOX=\vbox{}
\newbox\RefBoxOne   \setbox\RefBoxOne=\vbox{}

\expandafter\ifx\csname normal\endcsname\relax\def\normal{\null}\fi

\def\Eqno{\global\advance\EQNO by 1 \eqno(\the\EQNO)%
    \gdef\label##1{\xdef##1{\nobreak(\the\EQNO)}}}
\def\Fig#1{\global\advance\FIGNO by 1 Figure~\the\FIGNO%
    \global\setbox\FIGBOX=\vbox{\unvcopy\FIGBOX
      \narrower\smallskip\item{\bf Figure \the\FIGNO~~}#1}}
\def\Ref#1{\global\advance\REFNO by 1 \nobreak[\the\REFNO]%
    \global\setbox\REFBOX=\vbox{\unvcopy\REFBOX\normal
      \smallskip\item{\the\REFNO .~}#1}%
    \gdef\label##1{\xdef##1{\nobreak[\the\REFNO]}}}
\def\Section#1{\SUBSECNO=0\advance\SECNO by 1
    \bigskip\leftline{\bf \the\SECNO .\ #1}\nobreak}
\def\Subsection#1{\advance\SUBSECNO by 1
    \medskip\leftline{\bf \ifcase\SUBSECNO\or
    a\or b\or c\or d\or e\or f\or g\or h\or i\or j\or k\or l\or m\or n\fi
    )\ #1}\nobreak}
\def\Footnote#1{\global\advance\FOOTNO by 1
    \footnote{\nobreak$\>\!{}^{\the\FOOTNO}\>\!$}{#1}}
\def\SameFootnote{$\>\!{}^{\the\FOOTNO}\>\!$}

\def\References{\bigskip\centerline{\bf REFERENCES}
                \smallskip\copy\REFBOX}
\def\NewRefPage{\setbox\RefBoxOne=\vbox{\unvcopy\REFBOX}
		\setbox\REFBOX=\vbox{}
		\def\References{\bigskip\centerline{\bf REFERENCES}
                		\nobreak\smallskip\nobreak\copy\RefBoxOne
				\vfill\eject
				\smallskip\copy\REFBOX}
		\def\NewRefPage{}}




\font\tenbm=cmmib10
\font\ninei=cmmi9
\newfam\bmfam

\def\tenpointbmit{
\textfont\bmfam=\tenbm
\scriptfont\bmfam=\seveni
\scriptscriptfont\bmfam=\fivei
\def\bmit{\fam\bmfam\tenbm}
}

\tenpointbmit

\mathchardef\Gamma="7100
\mathchardef\Delta="7101
\mathchardef\Theta="7102
\mathchardef\Lambda="7103
\mathchardef\Xi="7104
\mathchardef\Pi="7105
\mathchardef\Sigma="7106
\mathchardef\Upsilon="7107
\mathchardef\Phi="7108
\mathchardef\Psi="7109
\mathchardef\Omega="710A
\mathchardef\alpha="710B
\mathchardef\beta="710C
\mathchardef\gamma="710D
\mathchardef\delta="710E
\mathchardef\epsilon="710F
\mathchardef\zeta="7110
\mathchardef\eta="7111
\mathchardef\theta="7112
\mathchardef\iota="7113
\mathchardef\kappa="7114
\mathchardef\lambda="7115
\mathchardef\mu="7116
\mathchardef\nu="7117
\mathchardef\xi="7118
\mathchardef\pi="7119
\mathchardef\rho="711A
\mathchardef\sigma="711B
\mathchardef\tau="711C
\mathchardef\upsilon="711D
\mathchardef\phi="711E
\mathchardef\cho="711F
\mathchardef\psi="7120
\mathchardef\omega="7121
\mathchardef\varepsilon="7122
\mathchardef\vartheta="7123
\mathchardef\varpi="7124
\mathchardef\varrho="7125
\mathchardef\varsigma="7126
\mathchardef\varphi="7127



%
%
\twelvepoint			
%
%



\font\tenbm=cmmib10
\font\ninei=cmmi9
\newfam\bmfam

\def\tenpointbmit{
\textfont\bmfam=\tenbm
\scriptfont\bmfam=\seveni
\scriptscriptfont\bmfam=\fivei
\def\bmit{\fam\bmfam\tenbm}
}

\tenpointbmit

\mathchardef\Gamma="7100
\mathchardef\Delta="7101
\mathchardef\Theta="7102
\mathchardef\Lambda="7103
\mathchardef\Xi="7104
\mathchardef\Pi="7105
\mathchardef\Sigma="7106
\mathchardef\Upsilon="7107
\mathchardef\Phi="7108
\mathchardef\Psi="7109
\mathchardef\Omega="710A
\mathchardef\alpha="710B
\mathchardef\beta="710C
\mathchardef\gamma="710D
\mathchardef\delta="710E
\mathchardef\epsilon="710F
\mathchardef\zeta="7110
\mathchardef\eta="7111
\mathchardef\theta="7112
\mathchardef\iota="7113
\mathchardef\kappa="7114
\mathchardef\lambda="7115
\mathchardef\mu="7116
\mathchardef\nu="7117
\mathchardef\xi="7118
\mathchardef\pi="7119
\mathchardef\rho="711A
\mathchardef\sigma="711B
\mathchardef\tau="711C
\mathchardef\upsilon="711D
\mathchardef\phi="711E
\mathchardef\cho="711F
\mathchardef\psi="7120
\mathchardef\omega="7121
\mathchardef\varepsilon="7122
\mathchardef\vartheta="7123
\mathchardef\varpi="7124
\mathchardef\varrho="7125
\mathchardef\varsigma="7126
\mathchardef\varphi="7127



\centerline{\bigbold SOLUTIONS FOR NEUTRAL}\vskip 0.7cm
\centerline{\bigbold AXI-DILATON GRAVITY IN 4-DIMENSIONS}
\bigskip\bigskip\bigskip

\centerline{T Dereli${}^{\dagger}$}
\medskip
\centerline{M \"Onder${}^{\dagger}{}^{\dagger}$}
\medskip
\centerline{Robin W Tucker}
\medskip

\centerline{\it School of Physics and Materials,}
\centerline{\it University of Lancaster,
		Bailrigg, Lancs. LA1 4YB, UK}
\centerline{\tt r.w.tucker{\rm @}lancaster.ac.uk}

\vskip 1cm
\vskip 2cm

\bigskip\bigskip\bigskip\bigskip

\centerline{\bf ABSTRACT}
\vskip 1cm

\midinsert
\narrower\narrower\noindent


We examine  a 1 parameter class of actions describing the gravitational
interaction between a pair of scalar fields and Einsteinian gravitation.
When the parameter is positive the theory corresponds to
an axi-dilatonic sector of low energy
string theory.
We exploit an  SL(2,R) symmetry of the theory to construct a family of
electromagnetically neutral solutions with non-zero
axion and dilaton charge from solutions
of the Brans-Dicke theory. We also comment on solutions to the theory with
negative coupling parameter.

\endinsert


\def\bff#1{{\bf #1}}
\def\br{\hfill\break}


\vfill

${}^{\dagger}$
{ Mathematics Department, Middle East Technical University, Ankara, Turkey}

${}^{\dagger}{}^{\dagger}$
{ Physics Department, Haceteppe  University, Ankara, Turkey}

\eject
\headline={\hss\rm -~\folio~- \hss}     

\Section{Introduction}

\def\pprime{^\prime}
\def\frac#1#2{{#1\over #2}}
\def\a{\alpha}
\def\b{\beta}

\def\wd{\wedge}

\def\ld{........}


\def\l{\lambda}

\def\KE#1{d\,#1\wedge *d\,#1}
\def\CKE#1{d\,#1\wedge * d\,#1^{*}}
\def\iCKE#1#2{d\,#1\wedge i_#2 * d\,#1^{*}}
\def\ttau#1#2{i_#1 d#2\wd * d#2 +d#2 \wd i_#1 * d#2}
\def\aa{{\cal A}}
\def\wd{\wedge}

Among the states of low energy string theory
one may identify a massless spin two graviton field in spacetime and
massless spin one fields describing the Yang-Mills interactions.
Besides the graviton,
there arise also an  antisymmetric tensor  and scalar (dilaton) field
with  specific couplings to each other and to the graviton
\Ref{E S Fradkin, A A Tseytlin, Nucl. Phys \bff {261} (1985) 1\br
M Dine, N Seiberg, Phys. Rev. Letts. \bff {55} (1985) 366\br
C G Callan, D Friedan, E J Martinec, M J Perry. Nucl. Phys. \bff {B262}
(1985) 593\br
}.
Depending on the topology of the compactification of internal dimensions
there appear further scalar excitations with well defined coupling schemes.
The pattern of all these couplings gives rise to  new kinds
 of ``duality'' symmetries
that may be used to discover new kinds of dilatonic black holes in
4-dimensions
\Ref{ A Shapere, S Trivedi, F Wilczek, Mod. Phys. Letts. \bff{A6} (9191)
2677}.
In the presence of the $U(1)$ gauge field such symmetries have been used to
discover electrically and magnetically charged axi-dilaton ``black-holes''
\Ref{G W Gibbons, Nucl. Phys \bff {207} (1982) 337\br
G W Gibbons,  K Maeda Nucl. Phys \bff {298} (1988) 741\br
D Garfinkle, G T Horowitz, A Strominger, Phys. Rev. \bff{D43} (1991) 3140}
\Ref{B A Campbell, M J Duncan, N Kaloper, K A Olive, Phys. Letts.
\bff{B251} (1990) 34\br
B A Campbell,  N Kaloper, K A Olive, Phys. Letts.
\bff{B263} (1991) 364\br
R Kallosh, T Ort\'in  Phys. Rev.
\bff{D48} (1992) 1006\br
S Mignemi, N R Stewert, Phys. Letts.
\bff{B298} (1993) 299\br
A Sen  Phys. Letts.
\bff{B303} (1993) 22
}
\Ref{ R Kallosh, A Linde, T Ort\'in  A Peet, A van Proeyen  Phys. Rev.
\bff{D46} (1992) 5278 \br
R Kallosh,  A Peet,   Phys. Rev.
\bff{D46} (1992) R5223 \br
T Ort\'in   Phys. Rev.
\bff{D47} (1993) 3136 \br
R Kallosh,  D Kastor, T Ort\'in, T Torma,  preprint (1994)}
Such solutions tend to flat spacetime when the electric and magnetic
charges tend to zero. It is of interest to enquire about the existence of
electromagnetically neutral solutions with non-trivial axion and dilaton
charge.

Since the above  symmetries
may be expressed either in terms of the geometry  of the
metric of the string graviton  or in terms of a Weyl scaled
metric (the so called Einstein frame) our discussion begins with
an action that exhibits  a manifest
SL(2,R) symmetry.

We examine an electrically neutral $U(1)$ sector of  theory
described  by the spacetime classical action $S[g,\zeta]=\int \Lambda$ where
$$\Lambda={\cal R} *1 -\eta\frac{\CKE\zeta}{q^2}.\Eqno$$\label\Actionone
Here ${\cal R}$ is the curvature scalar of the Levi-Civita connection
associated with the spacetime metric $g$ and in terms of the real scalar fields
$\aa$ and $q$ we define the complex scalar
$\zeta=\frac{\aa+iq}{2}$.
The coupling parameter $\eta$ is real and denotes the ratio of the
gravitational coupling constant to the scalar field interaction in the
action.
Expressed in terms of $\aa$ and $q$
$$\Lambda={{\cal R}} *1 -\frac{\eta}{4q^2}\KE q -\frac{\eta}{4q^2}\KE
\aa.\Eqno $$\label\Actiontwo
The action \Actionone\  is stationary under $g$ and $\zeta$ variations
if
$$R^{bc}\wd *(e_{a}\wd e_{b}\wd e_{c}) +\frac{\eta}{ q^2}Re(i_a\CKE\zeta
+\iCKE\zeta a ) =0\Eqno$$
$$\eta\frac{d * d\zeta}{q^2} +
\frac{2i\eta}{q^3}\KE\zeta =0\Eqno$$\label\zetaeqn
Taking the real and imaginary parts of \zetaeqn\  and writing in terms of
$\aa$ and $q$ these field equations become
$$R^{bc}\wd *(e_{a}\wd e_{b}\wd e_{c}) +\frac{\eta}{4q^2}\tau_a[q]+
\frac{\eta}{4q^2}\tau_a[\aa]=0\Eqno$$\label\eineqn
$$d * \frac{d\aa}{q^2}=0\Eqno $$\label\aaeqn
$$d * d q -\frac{\KE q}{q} + \frac{\KE{\aa}}{q}=0 \Eqno $$\label\qeqn
where the stress forms
$$\tau_a[q]=\ttau a q$$
$$\tau_a[\aa]=\ttau a \aa .$$
We are interested in stationary spherically stmmetric
solutions to these equations.

\Section{Relation to Low Energy String States}

The relation of \Actionone\  to electromagnetically neutral
axion-dilaton field theory predicted by string theory can be elucidated by
making a Weyl transformation on the Einstein metric:
${\bf{g}}=\lambda^2 g$
where  $\lambda$ is a positive scalar field.
If we denote the Hodge map associated with ${\bf{g}}$ as $\star$ then for
any p-form $\beta$ in  four dimensions one easily derives
$$*\beta=\l^{2p-4}\star\beta.\Eqno$$
Thus
$$\frac{\CKE\zeta}{q^2}=\frac{d\zeta \wd \star d \zeta^*}{\l^2 q^2}.\Eqno $$
{}From the definition of the curvature in terms of the metric the curvature
scalar associated with ${\bf{g}}$ may be expressed as
$${\cal R} *1=\frac{1}{\lambda^2}
{\bf { R}} \star 1 +\frac{6}{\l^4} d\l \wd \star d\l\quad\quad mod(d)\Eqno$$
If we choose
$\l^2=\frac{1}{q}$
then  the action form is expressed as
$$\Lambda= q{\bf { R}} \star 1 -\frac{\eta -6}{4q}dq \wd \star dq
 -\frac{\eta}{4q}d\aa \wd \star d \aa\Eqno $$
which should be compared with \Actiontwo.
For
$\eta > 0$ we may identify $\aa$ as the axion field and $\phi$ as the dilaton
of string theory where $ q=e^{-\phi}.$
In the absence of Yang-Mills gauge fields and higher order gravitational
and Yang-Mills Chern-Simons couplings we may express the action in terms
of the {\it closed} 3-form $H$
by writing
$d\aa=e^{-\phi} \star H.$ In terms of  $\phi$ and $H$ the effective
string action form  is then
$$\Lambda= e^{-\phi}\{{\bf { R}} \star 1 -
\a d\phi\wd \star d\phi +\b H\wd\star H\} \Eqno $$
where
$\a=\frac{\eta-6}{4}$  and $\b=\frac{\eta}{4}$ so $\b-\a=\frac{3}{2}.$

\Section{Consistency with Variations of the String Action}

Before seeking  solutions to \eineqn\ , \aaeqn\  and \qeqn\
 we must verify that such field equations
can be derived from the low energy string action expressed in terms of the
{\it closed} form $H$ in order to justify the claim that the theories
correspond under a Weyl scaling.
In order to accommodate the closure condition on $H$ we  vary the
action $S[g,q,H,\mu]=\int\Lambda\pprime$ where
$$\Lambda\pprime= {\cal R} *1 -\frac{\eta}{4 q^2}\KE q -\frac{\eta}{4}q^2 H\wd
* H +\mu dH .\Eqno$$\label\Lambdap
Since
$$\mu \,dH=H\wd d\mu \quad mod(d)\Eqno$$\label\dmu
the last term in \Lambdap\ imposes the closure constraint
\Ref{T Dereli, R W Tucker Class. Quan. Grav. \bff{4} (1987) 791}.
This action is stationary under variations of $ g,q,H$ and $\mu$ if
$$R^{bc}\wd *(e_{a}\wd e_{b}\wd e_{c}) +\frac{\eta}{4q^2}\tau_a[q]+
\frac{\eta q^2}{4}\tau_a[H]=0\Eqno$$
$$\frac{\eta}{2} d(\frac{*dq}{q^2}) + \frac{\eta}{2 q^3}\KE q
-\frac{\eta}{2} H\wd *H=0\Eqno$$
$$\frac{\eta}{2}q ^2 *H=d(\mu)\Eqno$$\label\closure
$$dH=0.\Eqno$$\label\Heqn
Introducing the axion field $\aa$ with the definition
$$q^2 *H=d\aa\Eqno$$
implies
$$d*(q^2H)=0\Eqno$$
consistent with \closure\
while \Heqn\  implies \aaeqn\ .
With the aid of the relations  
$$ \frac{\eta q^2}{4}\tau_a[H]\equiv\frac{\eta q^2}{4}\{i_aH\wd *H +H\wd
i_a*H\}$$
$$=\frac{\eta}{4 q^2}\ttau a \aa =\frac{\eta}{4 q^2}\tau_a[\aa]\Eqno$$
$$H\wd *H=-\frac{1}{q^4}\KE \aa$$
the field equations reduce to \eineqn\ \aaeqn\ and \qeqn\  as desired.
Furthermore using \dmu\ and $d(\mu)=\frac{\eta}{2}q ^2 *H$ in \Lambdap\ we find
$$\Lambda\pprime= {\cal R} *1 -\frac{\eta}{4 q^2}\KE q +\frac{\eta}{4}q^2 H\wd
* H  \quad\quad\quad mod(d)\Eqno$$
(note the sign of the third term  in \Lambdap). Substituting the relations
$\frac{d\aa}{q^2}=*H$ and $\frac{*d\aa}{q^2}=H$ in this we recover the
action form \Actionone\ .

\Section{Solutions for $\eta >0$}
\def\xmp{(\frac{\xi-1}{\xi+1})}
\def\xpm{(\frac{\xi+1}{\xi-1})}

The action defined by \Actionone\  is invariant under
$$g\mapsto g\Eqno$$
$$\zeta\mapsto \frac{c_1\zeta +c_2}{c_3\zeta +c_4}\Eqno$$\label\trans
where the  constants $c_j$ satisfy
$c_1c_4-c_2c_3=1.$
If one takes $\zeta$ real then the theory can be mapped by a Weyl
transformation  to the Brans-Dicke theory
\Ref{C H Brans, Phys. Rev. \bff{125} (1962) 2194\br
A I Janis, E T Newman, J Winicour, Phys. Rev. Letts. \bff {20} (1968) 878\br
A I Janis, D C Robinson J Winicour, Phys. Rev.  \bff {168} (1969) 1729}.
Consequently any solution of
that theory can be used as a representative of the class of solutions that
lie on the $SL(2,R)$ orbit of solutions solving \eineqn\ \aaeqn\  and
\qeqn . For $\eta>0$ such an orbit will then give a class of axi-dilatonic
solutions with zero electromagnetic charge.
Of particular interest is the
orbit described by  the metric
$$g=-\xpm^{2a_1} dt\otimes dt + (\xi^2 -1)^2 \xmp^{2a_1} g_3\Eqno$$
with dilaton
$$q=\xpm ^{a_2}\Eqno$$ and  axion $ \aa =0.$
Here we employ isotropic coordinates $t,r,\theta,\psi$ and $g_3$ denotes the
standard metric on ${\bf R}^3$ in terms of spherical polars $r,\theta,\psi.$
This is a particular solution if the real constants $a_1,a_2$ are
constrained by the relation
$a_1^2 +\eta a_2^2=1$ and $\xi\equiv -\frac{m}{2r}$
where $m$ is a real constant.
The case $a_2=0, a_1=1 $ corresponds to the Schwarzschild solution with
mass $m$.
Writing
$\phi =-\ln(q)$
we have for  $\|\xi\| << 1$
$$\phi=a_2\ln\xmp\simeq \frac{a_2 m}{r}\, + \,\ld\Eqno $$
and we identify the dilatonic charge of this solution to be $a_2 m$.
New solutions with a non-trivial axion field can now be generated with the
transformation \trans\ . In particular with
$c_1=\cos\Theta,\, c_2=\sin\Theta,\, c_3=-\sin\Theta, \,c_4=\cos\Theta$
one finds
$$q\pprime=\frac{q}{\cos^2\Theta +\sin^2\Theta(\frac{q}{2})^2}\Eqno$$
$$\aa\pprime=\frac{2\sin\Theta\cos\Theta((1-\frac{q}{2})^2)}
{\cos^2\Theta +\sin^2\Theta(\frac{q}{2})^2}\Eqno$$
and a corresponding
$$*H\pprime=\frac{1}{2}\sin 2\Theta\, d\phi .\Eqno$$
This solution has
dilaton charge  $a_2\, m \cos 2\Theta$
and axion charge  $a_2\, m \sin 2\Theta$.
The purely dilatonic $U(1)$ charged
solutions in references
\Ref{A G Agnese, M La Camera, Phys. Rev. {\bf D49} (1994) 2126},
\Ref{M Rakhmanov, Phys. Rev. {\bf D50} (1994) 5155}
lie on orbits that intersect this orbit of neutral axi-dilaton solutions
when the electric (and magnetic) charge  is zero.

\Section{Solutions for $\eta < 0$}

\def\exp#1{\,e^{-#1\frac{p_2}{r}}}

It is of
interest to note that for $\eta <0$ one may find a non-trivial  $SL(2,R)$
orbit of solutions based on the metric
$$g=-e^{-\frac{2p_1}{r}} \,dt\otimes dt + e^{\frac{2p_1}{r}}\,g_3\Eqno$$
with
$$q=e^{-\frac{p_2}{r}}\Eqno$$
$$\aa=0\Eqno$$
provided the real constants $p_1,\, p_2$ satisfy
$$p_1^2+\eta p_2^2=0.$$
This particular solution has dilatonic charge $p_2$.
Under \trans\  with
real $c_j$ one
generates the axion field
$${\aa\pprime}{}=
\frac{8 c_2c_4 +2c_1c_3\exp 2}{4 c_4^2 +c_3^2\exp 2}\Eqno$$
with axion charge
$$16p_2\, \frac{2 c_3 c_4}{(4c_4^2+c_3^2)^2}$$
and dilaton field
$${q\pprime}{}=\frac{4\exp {}}{4 c_4^2 +c_3^2\exp 2}\Eqno$$
with
dilaton charge
$$4p_2 \,\frac{c_3^2-c_4^2}{(4c_4^2+c_3^2)^2}$$
The corresponding $H$ field is
$$*H\pprime=-c_3c_4\, d\phi .\Eqno$$
An interesting feature of this solution is that the curvature invariant
$$*(R^{ab} \wd *R_{ab})
=\frac{2\,e^{-4p_1/r}\,p_1^2\,(12r^2-16 p_1 r +2 p_1^2)}{r^8}$$
is bounded in the region $r\geq 0$ with $p_1>0$.
\Section{Conclusion}
We have constructed a class of electromagnetically neutral
solutions to the  axi-dilatonic sector of low energy string theory
by exploting a series of Weyl transformations and a manifest $SL(2,R)$
symmetry of the theory.
The theory has been formulated in the ``Einstein frame'' with an arbitrary
coupling parameter and solutions  examined for both positive and negative
values of this coupling parameter.
We note that the
singularity structure of these solutions in terms of the Levi-Civita
curvature of the Einstein metric may not coincide with the corresponding
curvature associated with the metric scaled by the dilaton solution.

\Section{ Acknowledgment}

RWT and TD  are grateful to EPSRC for providing facilities at the 15-th UK
Institute for Theoretical High Energy Physicists   at the University of
Southampton where this work was begun and
to the Human Capital and Mobility Programme of the European Union for partial
support. TD is grateful
to the School of Physics and Materials,
University of Lancaster, UK  for its hospitality.

\vfill\eject

\References

\bye
\bye